\documentstyle[12pt,oldlfont,osa]{revtex}
\title{Absorption spectrum of a one-dimensional chain with Frenkel's
exciton
under diagonal disorder represented by hyperbolic defects}
\author{G.G.Kozlov}
\begin{document}
\maketitle
\begin{abstract}
 A method is proposed for calculating the absorption spectrum of
a long one-dimensional closed-into-a-ring chain with Frenkel's
exciton under diagonal disorder. This disorder is represented by
the hyperbolic singularities of atomic fission. These defects
are shown to lead to a wing in the exciton zone of a chain
without defects. The form of the wing does not depend on the
relative positions or number of defects and its value is
proportional to the sum of the amplitudes of the defects. The
proposed method uses only the continual approximation.
\end{abstract}
\section{Introduction}

Progress in the theory of translation-symmetric systems is
largely due to the possibility of using the Bloch function
tools. Switching to the Bloch representation makes it possible
to appreciably simplify and, sometimes, even to solve a variety
of problems in the physics of translation-symmetric systems. The
absence of a similar universal approach in the theory of systems
without translation symmetry causes many mathematical
difficulties in the analysis of even the simplest systems of
this kind. One such system is a model with the Frenkel
exciton-type excitations, i.e. a set of two level atoms bound by
an interaction capable of carring the excitation from one atom
to another. One usually explores only the one-exciton region of
energy spectrum, which corresponds to the presence of a single
excited atom. If the atoms are arranged symmetrically and have
different level splits, the system is called diagonally
disordered. Two problems of this kind -- the Dyson problem
\cite{Dyson} and the Lloyd problem \cite{Lloyd} -- are two of the very
few that allow an exact solution. Approximate methods (such as
the coherent potential approximation \cite{Lifshits}) were
developed for these systems. The quality criterion for these
methods is often a comparison with data from computer
experiments. However, it is not usually clear why the
approximate calculations agrees (if it does) with the numerical
calculation. Therefore, in the author's opinion, even abstract
models of translation-nonsymmetric systems are of interest if a
mathematically convincing solution scheme can be proposed for
them. In the present paper, such an abstract model is proposed.

We consider the problem of the Frenkel exciton in an infinitely
long chain of two- level atoms that has been closed into a ring.
The interatomic interaction and the atomic fission energy as a
function of the atom's coordinate are
 
\begin{equation}
w(z)=V\exp (-|{z\over R}|),
\end{equation}

and
 
\begin{equation}
\epsilon(z)=\sum\limits_r^N{a_r\over(z-R_r)}, a_r>0
\end{equation}

respectively, i.e. diagonal disorder is introdused in the form
of a random number $N$ of hyperbolic singularities (hereinafter,
hyperbolic defects) which are randomly placed at points $R_r$
and have positive amplitudes $a_r$ (of constatnt sign). In
present paper, we propose a method of calculating the absorption
spectrum of this chain at zero temperature.

The absorption spectrum of a defectless chain is known to be a
singlet. Therfore, changes in the spectrum due to defects are
easier to see than changes in the density of states. It is shown
herein that hyperbolic defects give rise to a wing in the
exciton zone of a defectless chain and that its shape does not
depend on the number or relative positions of the defects. Also
its size is proportional to the sum of the amplitudes of the
defects, $\sum\limits_r^N a_r$.  This result appears to be
somewhat unexpected. The calculation was carried out by the
method proposed in \cite{Kozlov}, which uses only the
continuation approximation (i.e. the replacement of the lattice
sums by integrals). The efficiency of this approach was
substantiated in \cite{Kozlov}

\section{Scheme of the method}

In oder to calculate the absorption spestrum of the chain
discribed in the Introduction, we use the method proposed in
\cite{Kozlov}. Here we give the results from that paper. If each
spectral line is thought of as a Lorentz line with width
$\delta$, then the absorption spectrum $A(\Omega)$ of a chain
with parameters (1),(2) and end coordinates $L$ and $-L$ can be
calculated by the formula

\begin{equation}
A(\Omega)=-{1\over\pi}\hbox{Im}
\int\limits_{-L}\limits^L{\rho\Psi(z)dz\over
E-\epsilon(z)}
\end{equation}

where $\Psi(z)$ is defined by

\begin{equation}
d^2\Psi/dz^2+\left({1\over R}\right)^2\left({W\over
E-\epsilon(z)}-1\right)\Psi=-\left({1\over R}\right)^2
\end{equation}

Here $R$ is the radius of interaction (1), $W=2\rho VR$ is the
width of the exciton zone of the defectless chain, $\rho$ is the
density of the atoms in the chain, $\epsilon$ is the atomic
fission energy, $\Omega$ is the energy of the incident light
quantum, and $E=\Omega+i\delta,$ $\delta>0$. For a chain that
closed into a ring, the function $\Psi(z)$ must satisfy the
cyclicity conditions
 
\begin{equation}
$$ \Psi(L)=\Psi(-L)$$
\end{equation}
\begin{equation}
$$d\Psi(L)/dz=d\Psi(-L)/dz$$
\end{equation}

Integrating (4) over $z$ from $L$ to $-L$ and taking into
account (5) and (6), we find that the integral in (3) can be
represented as follows:

\begin{equation}
2VR\int\limits_{-L}\limits^L{\rho\Psi(z)dz\over E-\epsilon(z)}=
\int\limits_{-L}\limits^L(\Psi(z)-1)dz
\end{equation}

Thus the task is redused to calculating the integral
$\int\limits_{-L}^L \Psi(z)dz$. The idea is to calculate this
integral not along the real axis but over semicircle of radius
$L$ in the upper half-plane of the complex variable $z$. This is
possible if $\Psi(z)$ does not have singularities (poles or
branch points) in the upper half-plane. Next, if $\epsilon(z)$
decreases in the above semicircle, we have the following
relationship: the lager the semicircle radius (chain lenth), the
more accurately Eq(4) can be solved by perturbation theory on
semicircle. As $L\rightarrow\infty$, the error tends to zero. It
is known \cite{Smirnov} that the solution to (4) has no
singularities in the upper half-plane of complex $z$ if the
function $W\over(E-\epsilon(z))$ is a single-valued and has no
poles there.

Let us show that this is the case for $\epsilon(z)$ of form (2),
where $E$ has a small positive imaginary part. If $E$ is real,
the equation $E-\epsilon(z)=0$ has $N$ real roots $z_k(E)$,
$k=1,...,N$. If $E$ has small imaginary part term $i\delta$,
$\delta>0$, the roots obtain an increment

$$\delta z_k={i\delta\over{d\epsilon(z_k)/dz}}$$

Since $\epsilon(z)$ defined by (2) is decreasing, all roots are
shifted to the lower half-plane and, thus, in our case,
$W\over(E-\epsilon(z))$ does not have poles in the upper
half-plane. In the next section, we calculate the spectrum by
discribed scheme.

\section{Calculation of the spectrum}

Let us represent (4) in the form

\begin{equation}
\Psi^{\prime\prime}+q^2\Psi+\Delta(z)\Psi=-(1/R)^2
\end{equation}

where

\begin{equation}
q^2\equiv\left({1\over R}\right)^2\left({W\over E}-1\right)
\end{equation}
\begin{equation}
\Delta(z)\equiv\left({1\over
R}\right)^2{W\epsilon(z)\over[E-\epsilon(z)]E}
\end{equation}

Here $q$ is the wave vector of the exciton with energy $E$ in
the defectless chain, the function $\Delta(z)$ describes the
defects and decreases in the semicircle of radius $L$ (below we
call it large semicircle) in the upper half-plane of complex
$z$. Let us find the solution of (8) in the form
$\Psi(z)=\Psi_0(z)+\Psi_1(z)+...$, where the lower index
indicates the order of correction $\Delta(z)$. These corrections
obey the following equations:

\begin{equation}
\Psi_0^{\prime\prime}+q^2\Psi_0=-(1/R)^2
\end{equation}
\begin{equation}
\Psi_1^{\prime\prime}+q^2\Psi_1=-\Delta\Psi_0
\end{equation}

It is not difficult to solve Eqs.(11) and (12). After that, in
the approximation under consideration, $\Psi(z)$ becomes
 
\begin{eqnarray}
\Psi(z)&=&-\left({1\over qR}\right)^2+\left({1\over qR}\right)^2
\int\limits_L^z
{\Delta(\xi)\over q}\sin[q(z-\xi)]d\xi+\\ &&+C_+\left(
e{^i}{^q}{^z}-
\int\limits_L^z{\Delta(\xi)\over
q}e{^i}{^q}{^\xi}\sin[q(z-\xi)]d\xi\right)+\nonumber\\&&+
C_-\left(e{^-}{^i}{^q}{^z}-\int\limits_L^z{\Delta(\xi)\over
q}e{^-}{^i}{^q}{^\xi}\sin[q(z-\xi)]d\xi\right)\nonumber
\end{eqnarray}

All integrations begin at the right end of the chain and go over
the large semicircle. Since $\Delta(z)$ is small there, (as
$L\rightarrow\infty$), we expect that solution (13) in the large
semicircle virtually coincides with the exact solution. To find
the constants $C_+$ and $C_-$, we use the solution (13) and
conditions (5) and (6), which is possible because the points
$\pm L$ lie in the large semicircle where this solution is
valid.  The result is the following set of equations for $C_+$
and $C_-$: 
    
\begin{equation}
C_+\{I_c(q,q)-2q\sin(qL)\}+C_-\{I_c(-q,q)-2q\sin(qL)\}=I_c(0,q)/(Rq)^2
\end{equation}
\begin{equation}
C_+\{iI_s(q,q)-2q\sin(qL)\}+C_-\{iI_s(-q,q)+2q\sin(qL)\}=iI_s(0,q)/(Rq)^2,
\end{equation}
 
where
 
\begin{eqnarray}
I_c(\alpha,\beta)\equiv\int\limits_L\limits^{-L}\Delta(\xi)e{^i}{^\alpha}{^\xi}
\cos[\beta(L+\xi)]d\xi\\&&
I_s(\alpha,\beta)\equiv-\int\limits_L^{-L}\Delta(\xi)e{^i}{^\alpha}{^\xi}
\sin[\beta(L+\xi)]d\xi
\end{eqnarray}

Recall that all integrations are carried out over the large
semicircle. We consider $q$ to be real and positive. This means
that the {\it \bf energy $E$ falls within the exciton zone of
defectless lattice}. Now we calculate integrals like (16),(17)
that appear in (14),(15). Wherever possible, we proceed to the
limit $L\rightarrow\infty$. Let us explain, for example, how 

$$I_s(0,q)=-\int\limits_L^{-L}d\xi\Delta(\xi)\sin[q(L+\xi)]$$

is calculated. In the large semicircle, only the increasing
exponent $\exp(-iq(L+\xi))$ should be retained in sine function.
On the other hand, this exponent tends to zero in the lower
large semicircle. Therefore,

$\int\limits_L^{-L}\Delta(\xi)\exp[-iq(L+\xi)]d\xi$

can be calculated over the entire large circle, which is quite
easy to do with the help of residues. We give the results of
calculating the integrals appearing in (14), (15) below:

\begin{eqnarray}
I_s(0,q)=\pi e^{-iqL}\sum\limits_\xi
\hbox{Res}[\Delta(\xi)e^{-iq\xi}]\\&& I_s(q,q)={1\over 2i}
e^{-iqL}\int\limits_L^{-L}\Delta(\xi)d\xi\\&& I_c(0,q)=\pi i
e^{-iqL}\sum\limits_\xi\hbox{Res}[\Delta(\xi)e^{-iq\xi}]\\&&
I_c(q,q)={1\over 2} e^{-iqL}\int\limits_L^{-L}\Delta(\xi)d\xi
\end{eqnarray}

Since $I_c(0,q)=iI_s(0,q)$ and $iI_s(q,q)=I_c(q,q)$, then
$C_-=0$ and we do not need the remainig integrals from (14),
(15). For $C_+$, we obtain the following formula:

\begin{equation}
C_+=\left({1\over Rq}\right)^2 {{2\pi i e^{-iqL}\sum\limits_\xi
\hbox{Res}[\Delta(\xi) e^{-iq\xi}]}\over
e^{-iqL}\int\limits_L^{-L}\Delta(\xi)d\xi-4q\sin(qL)}
\end{equation}

For the final calculation of $\int\limits_L^{-L}(\Psi(z)-1)dz$,
we also need the following integrals that appeared in (13):

$$\int\limits_L^{-L}dz\int\limits_L^z
d\xi\Delta(\xi)\sin[q(z-\xi)]=-{i\pi\over
q}e^{-iqL}\sum\limits_\xi\hbox{Res}[\Delta(\xi)
e^{-iq\xi}]+{1\over q}\int\limits_L^{-L} \Delta(z)dz$$
$$\int\limits_L^{-L}dz\int\limits_L^z
d\xi\Delta(\xi)e^{iq\xi}\sin[q(z-\xi)]=-{e^{-iqL}\over
2q}\int\limits_L^{-L} \Delta(z)dz$$

These expressions are derived by expanding the sine into
components, integrating by parts, retaining only the exponents
that increase in the large semicircle, and proceeding to
integration over the hole large circle as when deriving
(18)--(21). Taking into account that
 
$$\int\limits_L^{-L}{\epsilon(z)dz\over
E-\epsilon(z)}={i\pi\over E}\sum\limits_r a_r,$$

(integration goes over the large semicircle), it is easy to
obtain the following final formula for the absorption spectrum:

\begin{equation}
A(\Omega)=-{1\over\pi}\hbox{Im}\left({2L\rho\over
E-W}-i\pi\rho\left({1\over E-W}\right)^2\sum\limits_ra_r\right)
\end{equation}
 
$$E=\Omega+i\delta$$

\section{Discussion}

The first term in (23) yields a singlet and does not depend on
the presence of defects. The second term yields a wing in the
exciton zone. It is somewhat unexpected that its shape does not
depend on the relative positions $R_r$ of the defects. The wing
amplitude turns out to be proportional to the sum of the
amplitudes of the defects; this calculation has a "physical"
level of rigor. For a rigorous calculation , one should write
the entire perturbation series in $\Delta$ for the solution of
(4) and make sure that the influence of all terms of this
series, except the above-considered first-order term, tends to
zero in the large semicircle as $L\rightarrow \infty$. This was
done and result (23) turned out to be valid only for real $q$,
i.e., for energies falling within the excitone zone of the
defectless chain. For other energy values, the contribution from
terms of orders higher than 1 can not be ignored. This behavior
should be discussed.

As shown in \cite{Kozlov}, $\Psi(z)$ is related to the Green's
function of the system and, thus, solution (23) could be
analytically extended to the range of energies beyond excitone
zone of the defectless chain. This would be correct if we were
dealing with exact Green's function. However, Eq.(4) is derived
in \cite{Kozlov} in the continual approximation.  Therefore,
(23) can yield noticeable deviations from the exact spectrum
beyond the excitone zone, where the calculation of the previous
section is invalid. Computer analysis confirms this remark.

Figure 1 (there is no figure in this version of the paper, but
it is easy to imagine) displays the logarithmic absorption
spectra of two chains with different numbers, positions and
amplitudes of the hyperbolic defects at
$W=1,\rho=1,\delta=0.02,R=2$. In all cases the number of
particles was 600, i.e., $2L=600$. The upper part $a$ shows the
logarithmic absorption of a chain with five hyperbolic defects.
Their positions (numbers of the sites at which atomic fission
becomes infinite, i.e., these atoms are effectively absent) are
$200, 250, 300, 350, 400$. The respective amplitudes of the
defects are $2, 3, 3, 6, 1$. The lower part $b$ shows the same
for a chain with two defects having the amplitudes -- $5$ and
$5$ and positions -- $250$ and $350$. For each numeric spectrum,
a spectrum calculated by (23) (smoothed curves) is given. No
fitting (either in amplitude or in shape) was carried out. The
energy plotted in units of $W$. It is evidend from Fig.1 that
for $0<\Omega< 1$ (the excitone zone), the numerical spectra
almost completely coincide with the theoretical spectra and the
shape of the spectra does not depend on the size or positions of
the defects. The discrepancy for $\Omega\approx1$ is due to the
finiteness of the chain because, in this energy range, $q^2
(q^2\approx 0)$ can be comparable to $\Delta(z)$, which is
finite in the large semicircle due to the finiteness of the
chain. In the region beyond the exciton zone, only agreement "in
the mean" is possible (it can be improved by increasing the
imaginary part of the energy $\delta$). In this region, the
spectrum does depend on the positions and sizes of the defects.
It may be possible to apply the proposed approach to more
realistic models of chains with defects.

\end{document}